\documentclass[preprint,12pt]{elsarticle}

\usepackage{amssymb}

\usepackage{amsmath}
\usepackage[dvipsnames]{xcolor}
\usepackage{ulem}
\usepackage{array}

\begin{document}

\begin{frontmatter}

\title{The performance of the TA$\times$4 surface detector array: 4.3 years of the first-half expansion}

\author[1]{R.U.~Abbasi}
\author[1,2]{T.~Abu-Zayyad}
\author[2]{M.~Allen}
\author[2]{J.W.~Belz}
\author[2]{D.R.~Bergman}
\author[3]{F.~Bradfield}
\author[2]{I.~Buckland}
\author[2]{W.~Campbell}
\author[4]{B.G.~Cheon}
\author[3]{K.~Endo}
% ...

\cortext[cor1]{Corresponding authors.}
\fntext[PurpleMountainObs]{Presently at: Purple Mountain Observatory, Nanjing 210023, China}
\fntext[BrookhavenNL]{Presently at: Physics Department, Brookhaven National Laboratory, Upton, NY 11973, USA}
\fntext[KoreaInstGMRD]{Presently at: Korea Institute of Geoscience and Mineral Resources, Daejeon, 34132, Korea}
\fntext[NASA_MSFC]{Presently at: NASA Marshall Space Flight Center, Huntsville, Alabama 35812, USA}

\affiliation[1]{
  organization={Department of Physics, Loyola University-Chicago},
  city={Chicago},
  state={Illinois},
  postcode={60660},
  country={USA}
}
\affiliation[2]{
  organization={High Energy Astrophysics Institute and Department of Physics and Astronomy, University of Utah},
  city={Salt Lake City},
  state={Utah},
  postcode={84112-0830},
  country={USA}
}
\affiliation[3]{
  organization={Graduate School of Science, Osaka Metropolitan University},
  city={Osaka},
  state={Osaka},
  postcode={558-8585},
  country={Japan}
}
\affiliation[4]{
  organization={Department of Physics and The Research Institute of Natural Science, Hanyang University},
  city={Seoul},
  state={Seoul},
  postcode={426-791},
  country={Korea}
}
% ...

\begin{abstract}
The Telescope Array (TA) experiment aims to reveal the origin of ultra-high-energy cosmic rays (UHECRs) by observing air showers using surface detectors (SDs), 
which spread over an area of approximately 700 km$^2$, and fluorescence detectors (FDs) viewing the skies above the SD array. 
The TA experiment has been observing UHECRs since 2008, and has reported an indication of clustering in the arrival directions of cosmic-ray events with energy greater than 57 EeV. 
To improve the exposure for anisotropy studies of UHECRs, the TA$\times$4 upgrade was designed to expand the observational area by approximately 2,000 km$^2$ with 500 additional SDs. 
Half of the planned upgrade, consisting of 257 SDs, was completed, and the newly installed array began operation in 2019. 
In addition to the expanded SD array, two FD stations were constructed for the TA$\times$4 experiment. 
In this paper, we present a study of the performance of the expanded SD array, including the energy resolution, angular resolution, 
and effective aperture, over the first 4.3 years of data acquisition. 
While the effective aperture varied initially due to changing detector states, it has stabilized since June 2023 with more than 90\% operational SDs.
Furthermore, a new inter-tower trigger system was implemented to connect six new communication towers to form two geographically separated arrays, increasing the effective aperture.
The time variation of this effective aperture, the resulting total exposure of approximately 3,500 km$^2$~sr~yr, and a comparison with the original TA SD array are presented to demonstrate the performance of the expanded array.
\end{abstract}




\end{frontmatter}


\section{Introduction}
\label{sec:int}
The Telescope Array (TA) experiment \citep{TelescopeArray:2008toq} is designed to observe ultra-high-energy cosmic rays (UHECRs) and, in part, to reveal their origins. 
The experiment is located in Utah, USA, at $39.3^{\circ}$\,N, $112.9^{\circ}$\,W, at an altitude of 1370 m above sea level. 
TA began observation in May 2008 with 507 surface detectors (SDs) which are spread over an area of about 700 km$^{2}$ area \citep{TelescopeArray:2012uws}, and with three fluorescence detector (FD) stations \citep{Tokuno:2012mi}. 
The SDs are arranged in a square grid pattern with 1.2 km spacing.

The TA experiment has reported an indication of a clustering of UHECRs with energies above 57 EeV (``TA hotspot'') in the first five years of data, 
within a 20$^{\circ}$ circle centered at R.A. = 146.7$^{\circ}$, Dec. = +43.2$^{\circ}$ (J2000) \citep{TelescopeArray:2014tsd}. 
The chance probability to observe the excess somewhere in the field of view is $3.7\times10^{-4}$, corresponding to a significance of 3.4$\sigma$. 
This excess may provide a clue to the origin of UHECRs. 
While no significant excess has been reported in the same region by the Pierre Auger Observatory \citep{PierreAuger:2024hrj}, 
the excess persists in TA data and the origin of the TA hotspot remains under discussion.

Motivated by the need for greater exposure to study the highest-energy cosmic rays, including anisotropy searches, 
TA$\times$4 has been designed to substantially increase the sample of UHECRs, especially for events with energies greater than 57 EeV.
The complete upgrade will quadruple the observational area of the original TA experiment by adding new SD arrays. 
As the first phase of this project, the deployment of 257 SDs was completed in March 2019, and observation with this half-expanded array began in April 2019.

In this paper, we present the performance of the half-expanded TA$\times$4 SD array, over the initial 4.3 years of data acquisition. 
The configuration of the TA$\times$4 SD array and the design of the individual detectors are summarized in Sec.~\ref{sec:daq}. 
The data acquisition and the operational status of the TA$\times$4 SD array are described in Sec.~\ref{sec: Trigger system} and Sec.~\ref{sec:workingstatus}, respectively. 
In Sec.~\ref{sec:performance}, the energy resolution, angular resolution, and effective aperture of the TA$\times$4 SD array are evaluated using Monte Carlo (MC) simulation. 
The last section, Sec.~\ref{sec:summary}, summarizes the paper.

\section{TA$\times$4 surface detector array}
\label{sec:daq}
\subsection{Array configuration}
\begin{figure}[!htbp]
\begin{center}
\includegraphics[width=95mm]{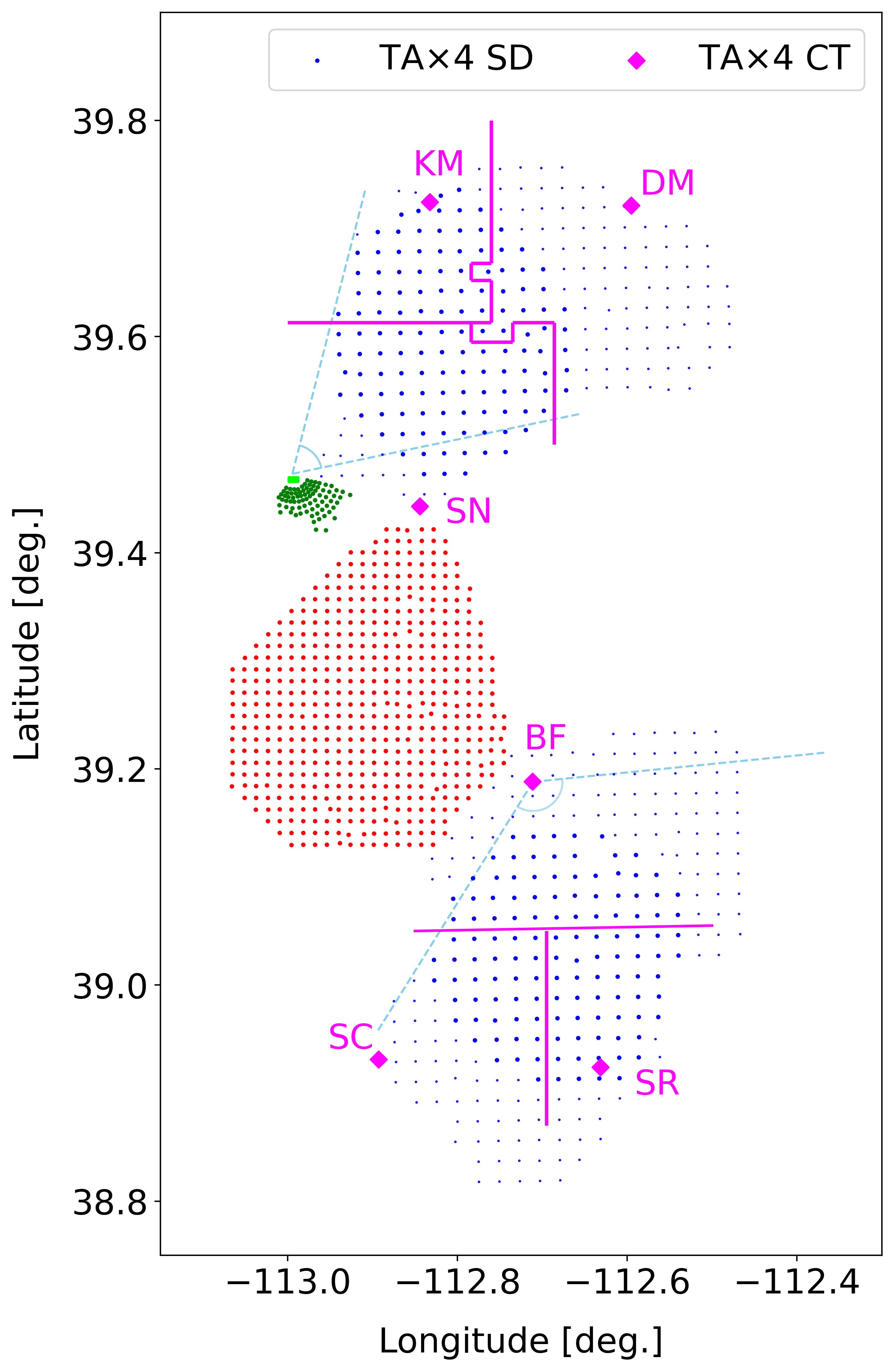}
\caption{The layout of the TA$\times$4 experiment. The red circles represent 507 TA SDs (TA SD array). 
The blue circles show the 257 newly deployed SDs for the TA$\times$4 upgrade in 2019 (TA$\times$4 SD northern/southern array). 
The blue dots show the sites still to be deployed in the future. 
The green circles represent 80 SDs deployed for the Telescope Array Low-energy Extension (TALE), 
and the light green square indicates the observational area of the TALE infill SD array, which consists of 50 SDs in a square grid with 100-m spacing. 
The pink diamonds show the locations of communication towers for the six TA$\times$4 sub-arrays. 
The pink solid lines mark the boundary of the sub-arrays. 
The fields of view of the TA$\times$4 FDs in azimuth are indicated by the dashed light blue lines.}
\label{fig:layout}
\end{center}
\end{figure}

The 257 newly deployed SDs are geographically divided into two regions, 
located to the north-east (the TA$\times$4 SD northern array, 130 SDs) and south-east (the TA$\times$4 SD southern array, 127 SDs) of the original TA SD array. 
The geometry of the TA$\times$4 extension was determined taking into account the availability of suitable land for long-term detector deployment, 
together with practical constraints such as land-use permissions, site accessibility, terrain, and communication infrastructure. 
Under these constraints, the northeastern and southeastern regions provided sufficiently large areas suitable for deployment. 
Figure~\ref{fig:layout} shows the configuration of the TA$\times$4 SD array. 
This half-expanded setup covers a combined geometric area of approximately 1,000 km$^{2}$.
While the original TA SDs are arranged with a 1.2 km spacing, the TA$\times$4 SDs feature an increased detector spacing of 2.08 km to optimize the effective area for higher-energy cosmic rays. 
This wider spacing is expected to raise the energy threshold for the new arrays to approximately $10^{19.8}$ eV to achieve 97\% trigger efficiency \citep{TelescopeArray:2021dri}, 
whereas the corresponding threshold for the original TA SD array is $10^{19.0}$ eV \citep{TelescopeArray:2012uws}.
The TA$\times$4 northern and southern arrays are each divided into three sub-arrays: Keg Mountain (KM), Desert Mountain (DM) and Smelter Knolls North (SN) for the northern array, 
Black Rock Mesa FD (BF), South Cricket (SC) and Sand Ridge (SR) for the southern array. 
The data acquisition for each sub-array is controlled by the electronics in the communication tower of the same name. 
Each sub-array initially took data independently; however, they were later linked together using a newly developed central trigger system, 
which is described in detail in Sec.~\ref{sec: Central trigger} and Sec.~\ref{sec:workingstatus}.

To simultaneously observe air showers above the TA$\times$4 SD arrays, 
four fluorescence telescopes at the Middle Drum and eight fluorescence telescopes at Black Rock Mesa FD station 
have been operational since January 2018 and June 2020, respectively. 

\subsection{Detector design}
\label{sec:TAx4SD}
The basic design of the newly deployed TA$\times$4 SDs is essentially the same as that originally deployed in the TA SD array, 
though with several significant internal upgrades (such as improved PMT linearity and modified electronics) as described in detail in \citep{TelescopeArray:2021dri}. 
The TA$\times$4 SD consists of two layers of plastic scintillators, each with dimensions of 150 cm in length, 200 cm in width, and 1.2 cm in thickness. 
The scintillation light of each layer is collected by its own photomultiplier tube (PMT) via wavelength-shifting fibers. 
Signals from each PMT are digitized by a 12-bit flash analog-to-digital converter (FADC) at a 50 MHz sampling rate.
Each SD communicates with the communication tower of the sub-array via a 2.4-GHz wireless link using a parabolic antenna. 
During daytime, a 120 W solar panel charges a 100 Ah battery, 
which is the sole source of power for each TA$\times$4 SD. 
One day of solar charging provides enough power to operate an SD for about one week.

\section{Data acquisition}
\label{sec: Trigger system}
The readout electronics for each SD records signals from the two scintillator panels when specific triggers are issued. 
There are three levels of triggers: \textit{Level-0} and \textit{Level-1} triggers are formed individually by each SD according to the particle flux detected by the scintillators. 
A \textit{Level-2} trigger is issued by the communication towers based on temporal coincidence and spatial adjacency. 
In this section, we describe the trigger systems and the recorded data. 
\subsection{Level-0 and Level-1 triggers}
\label{sec: Level-0 trigger}
A Level-0 trigger is issued when both scintillator layers of an SD record more than 15 FADC counts within 160 ns, 
which corresponds to approximately 0.3 minimum ionizing particles (MIPs). 
When this happens, the SD readout electronics save the FADC traces (``waveforms'') within 2.56 $\mu$s to a local synchronous dynamic random-access memory along with the trigger time. 
The typical rate of Level-0 triggers is approximately 750 Hz. 

\label{sec: Level-1 trigger}
The criteria for Level-1 trigger are the same as those of Level-0 trigger except for an increased FADC threshold of 150 FADC counts, 
which corresponds to approximately 3 MIPs. 
The time stamps of Level-1 triggers are sent to the assigned communication tower every second to form a Level-2 trigger, 
as described below. 
The typical rate of Level-1 triggers is approximately 20 Hz. 

\subsection{Level-2 trigger}
\label{sec: Level-2 trigger}
\begin{figure}
\centering
\includegraphics[width=16cm,clip]{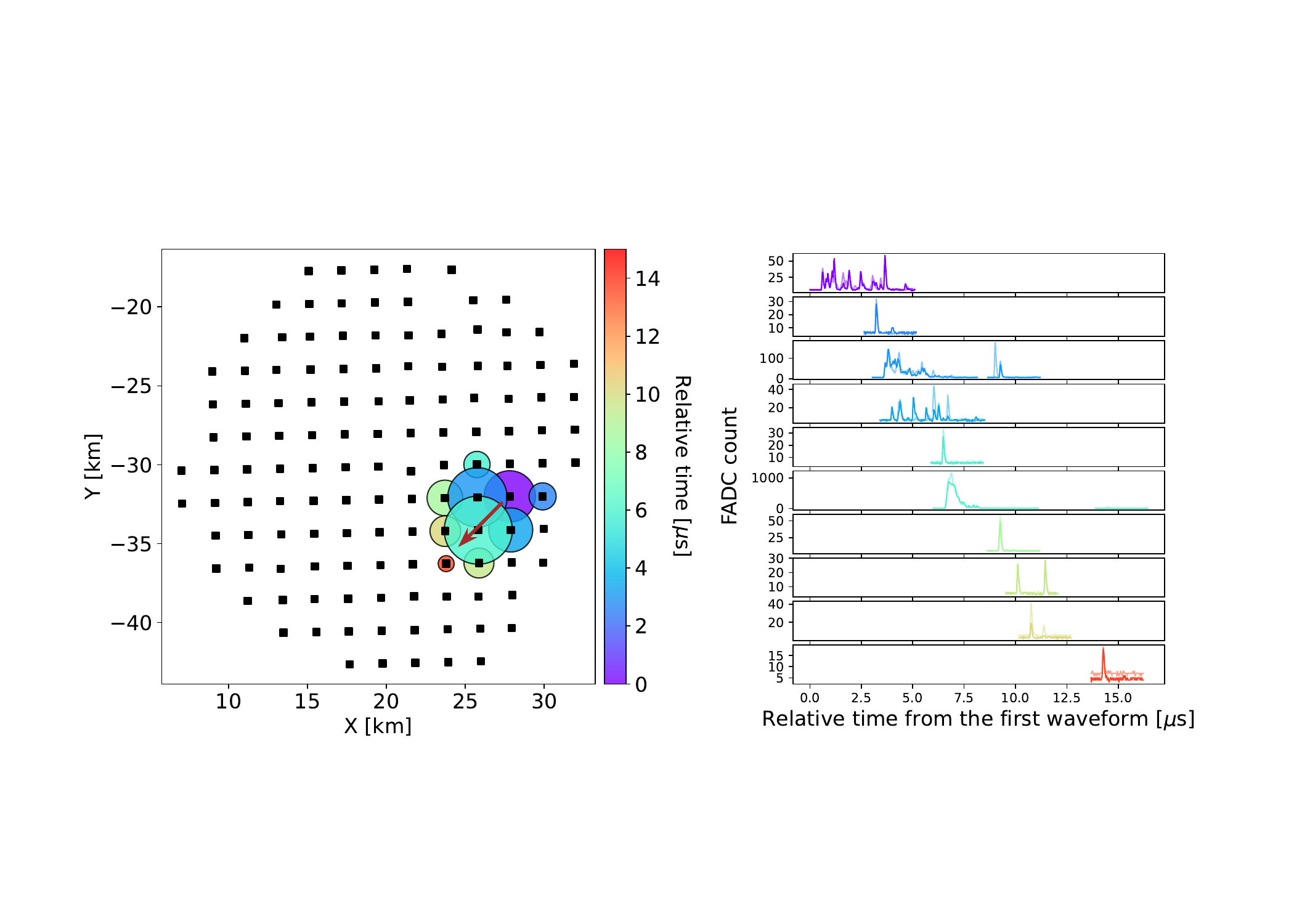}
\caption{An example of an air shower event. The primary energy and zenith angle are estimated to be $10^{19.8}$ eV and $36^{\circ}$, respectively. 
(Left) The footprint of an air shower event, observed by the TA$\times$4 SD southern array on 6 November 2022. 
The black squares represent SDs. 
The circles indicate positions of the SDs with saved waveforms. 
The area of each circle is proportional to $\log_{10}(\rho)$, where $\rho$ is the signal density measured by the corresponding SD. 
The relative signal time is indicated by the color of the circle. The arrow indicates the shower-propagating direction. 
(Right) The waveforms of the air shower event as measured by each SD. 
The color of each waveform represents the timing order, as displayed by circles in the left panel. 
The thick (thin) lines represent the upper-layer (lower-layer) waveforms.
The detailed definitions of the signal timing and the signal density are given in the main text of Sec.~\ref{sec: Level-2 trigger}.}
\label{fig: event-waveforms}
\end{figure}

The TA$\times$4 SD array has two types of Level-2 trigger systems. 
One is the \textit{SD self-triggers} including both \textit{sub-array triggers} (Sec.~\ref{sec: SD-self trigger}) and \textit{inter-tower triggers} (Sec.~\ref{sec: Central trigger}). 
The SD self-trigger is generated by the communication tower using only SD information. 
The other one is the so-called \textit{hybrid trigger}, described in Sec.~\ref{sec: hybrid trigger}, which is generated by FD stations. 
When a Level-2 trigger is issued, the waveforms saved locally at the SDs by Level-0 triggers are sent to the corresponding communication tower. 
A set of waveforms from coincident and adjacent SDs constitutes an air shower event and is saved in data storage at the tower. 
Figure~\ref{fig: event-waveforms} shows an example of an air shower event. 
To extract signal timing from these varied pulse profiles, 
the timing is defined as the first bin of the first four consecutive time bins in which the FADC count (after pedestal substraction) 
exceeds five times the standard deviation of the pedestal noise.
Additionally, the signal density $\rho$ (VEM/m$^2$) is constructed 
by converting the integrateed FADC counts into units of vertical equivalent muons (VEM) using the 1-MIP calibration histogram 
(detailed in Sec.~\ref{sec:calibdata}) and dividing by the 3 m$^2$ detector area.

\subsubsection{Sub-array trigger}
\label{sec: SD-self trigger}
All SDs within a TA$\times$4 sub-array communicate directly with a single communication tower. 
A Level-2 trigger occurs when any set of three adjacent SDs in a sub-array issues a Level-1 trigger within a time window of 14 $\mu$s. 
In this case, all waveforms locally saved by Level-0 triggers occurring within $\pm$32 $\mu$s of the Level-2 trigger time in the sub-array are collected by the tower from the detectors. 
This trigger method is the same as that used in the TA SD array except for the time window, which is 8 $\mu$s for the TA SD array. 
The increased time window is required due to the expanded detector spacing of TA$\times$4. 
Before the implementation of the inter-tower trigger, the formation of Level-2 triggers and subsequent waveform collection were independently performed in the six sub-arrays. 

\subsubsection{Inter-tower trigger}
\label{sec: Central trigger}
\begin{figure}[!htbp]
\begin{center}
\includegraphics[width=150mm]{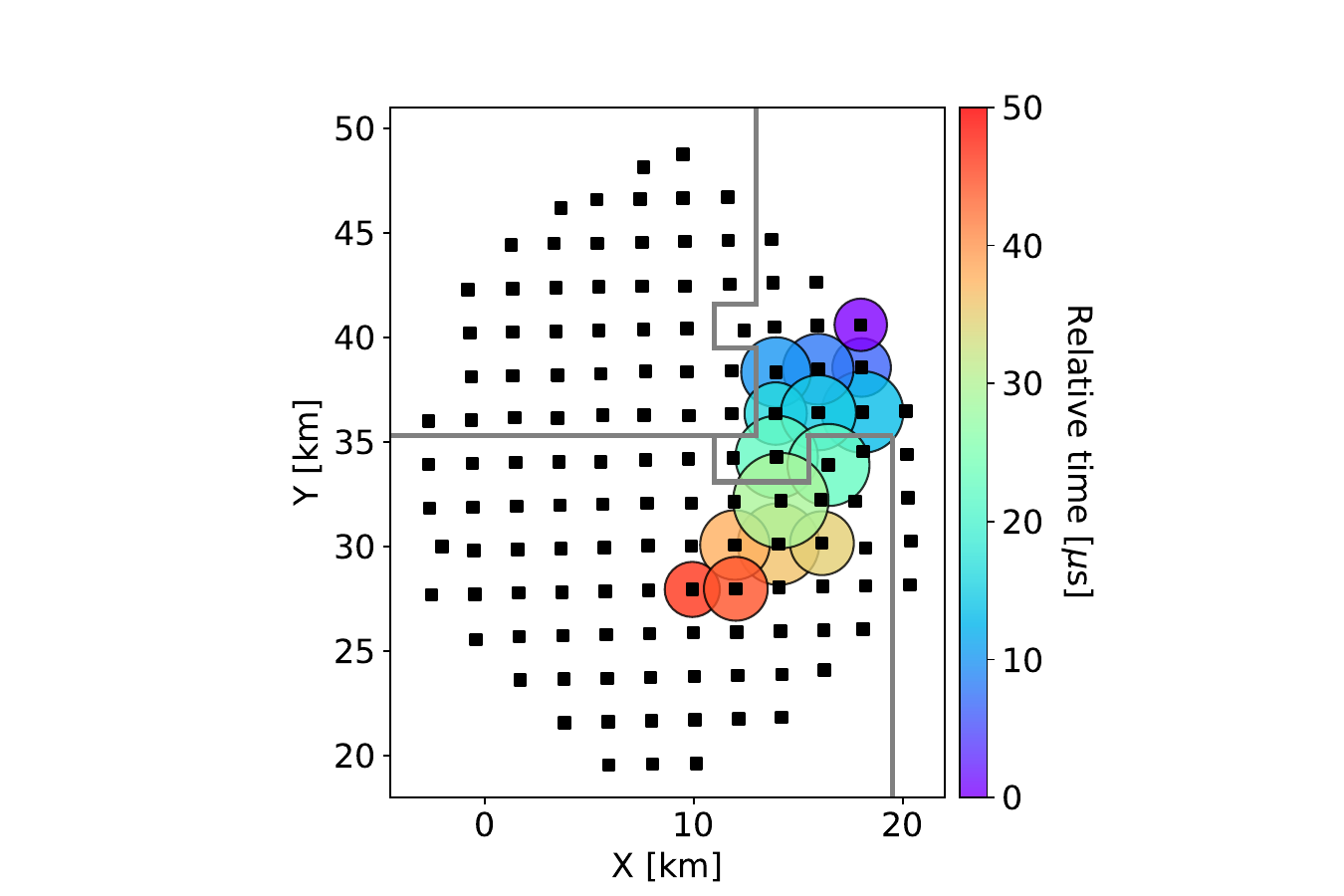}
\caption{An example of an event recorded by the inter-tower trigger in the TA$\times$4 northern SD array, 
observed on 27 January 2024. The zenith angle is estimated to be 80$^{\circ}$. 
The thick gray lines represent the boundaries of sub-arrays.}
\label{fig:ex_bd}
\end{center}
\end{figure}

The formation of Level-2 triggers for events with adjacent SDs that cross between sub-arrays requires the transfer of Level-1 trigger time stamps between communication towers. 
This system uses dedicated 5 GHz wireless links between towers~\citep{comtower}, along with an additional software layer to identify, collect, and build cross-boundary coincident waveforms into events. 
The available bandwidth of the inter-tower link is approximately several tens of Mbps, while the data size of a Level-1 trigger time stamp in each TA$\times$4 SD array is only several tens of kbps.
Therefore, we implemented a simple data transfer of all Level-1 trigger times from non-central towers to the central tower (\textit{i.e.} one nearest to the main TA SD array) every second for the inter-tower triggers.
The inter-tower trigger is then formed according to the following algorithm.
\begin{enumerate}
\item All Level-1 trigger times from KM (SR) and DM (SC) are sent to the central tower SN (BF) every second for the northern (southern) array.
\item Level-2 trigger conditions are identified from Level-1 trigger times over the full northern (southern) array.
\item If there are additional Level-2 triggers to the current Level-2 triggers at the central tower, requests to save waveforms are sent to all the towers.
\item Duplicate save requests are identified and discarded at each tower. When no Level-0 triggers are found in the time window, no save request is generated. 
\end{enumerate}
We show an example of an event triggered by the inter-tower trigger in Fig.~\ref{fig:ex_bd}.

\subsubsection{Hybrid trigger}
\label{sec: hybrid trigger}
In addition to the SD self-triggering system, an external trigger system from the FD stations, called a hybrid trigger, has also been implemented. 
When the FD station is triggered by a track-like air-shower event, partially reconstructed in real-time as originating from the direction of the TA$\times$4 SD array, 
the FD trigger time is sent to the communication towers by the hybrid trigger, and waveforms of SDs within $\pm64$ $\mu$s of the trigger time are collected by the communication tower. 
In particular, the hybrid trigger allows the identification and recording of air shower events at energies insufficient to issue an SD self-trigger \citep{MPotts_PhD2022}. 
The maximum rate of the TA$\times$4 hybrid trigger is 10 mHz.

\begin{figure}
\centering
\includegraphics[width=15cm,clip]{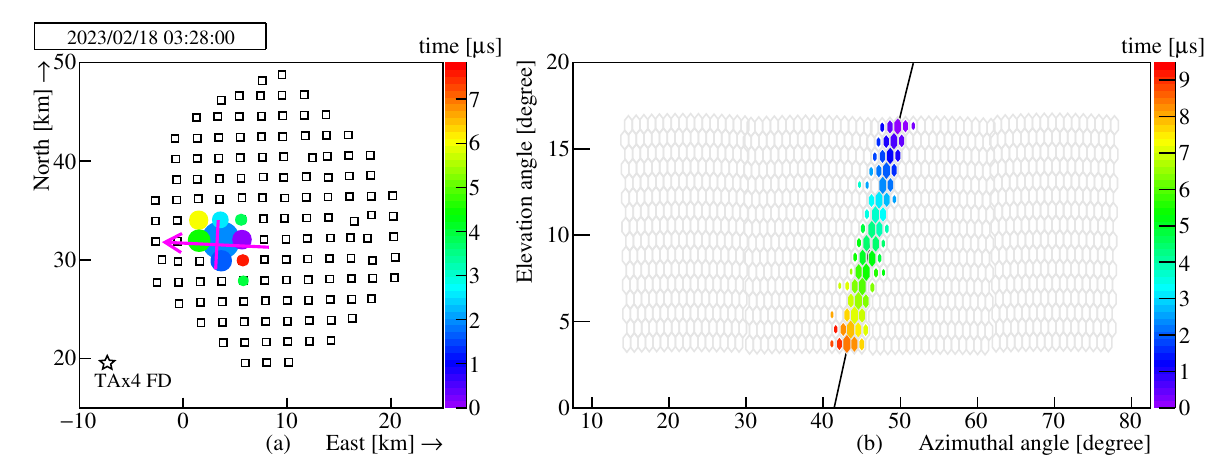}
\caption{An example of the hybrid-triggered event in the TA$\times$4
northern array on 18 February 2023. 
(Left) The SD shower footprint. 
(Right) The shower track as observed by the FD station.}
\label{fig:ex_hyb} 
\end{figure}

An example of a hybrid-triggered event is shown in Fig.~\ref{fig:ex_hyb}. 
The zenith and azimuthal angles reconstructed using only the SD array are 30.3${^\circ}$ and 101.2${^\circ}$, respectively, 
and those reconstructed using the combined FD and SD data are 30.9${^\circ}$ and 99.6${^\circ}$, respectively. 
The opening angle between the two reconstructed directions is 1.0${^\circ}$. 
Detailed descriptions of the hybrid trigger can be found in \citep{TelescopeArray:2021oqr}.

\subsection{Calibration data}
\label{sec:Lowleveldata}
\label{sec:calibdata}
The calibration constants used to convert from the integrated FADC counts to the number of VEM from the different SDs are extracted from the so-called \textit{1-MIP histograms}, the distribution of integrated FADC traces of an SD. 
When a Level-0 trigger is issued, the integral of the FADC counts over 12 time bins (from 80 ns before to 160 ns after the trigger) is accumulated into the 1-MIP histograms, 
which are collected at ten-minute intervals. 
By modeling the distribution of integrated FADC signals from single charged particles, we obtain a conversion factor from FADC count to energy deposit in units of MIP. 
The PMT gain was adjusted so that typically 50 FADC counts correspond to one MIP. 
Each SD also records a pedestal histogram by accumulating the FADC counts every 8 time bins (160 ns). 
The pedestal histogram is used to determine the baseline of the signal. 
When we reconstruct observed data, the corresponding values are used as the one MIP value and pedestal level. 
These values are reproduced during the actual data-taking periods in the Monte Carlo simulation described in Sec.~\ref{sec:MC} in order to reproduce actual SD data taking conditions.

In addition to the 1-MIP and pedestal histograms, each SD also records GPS information, trigger rate, battery condition, communication status, etc. 
These monitoring data are acquired every 10 minutes, allowing us to track the operational status of the SDs in detail and remove malfunctioning SDs. 
An example of such a malfunctioning SD is one with no GPS clock synchronization, which is then excluded from the reconstruction of air shower events. 

\section{Operational status}
\label{sec:workingstatus}

\begin{figure}[!htbp]
\begin{center}
\includegraphics[width=150mm]{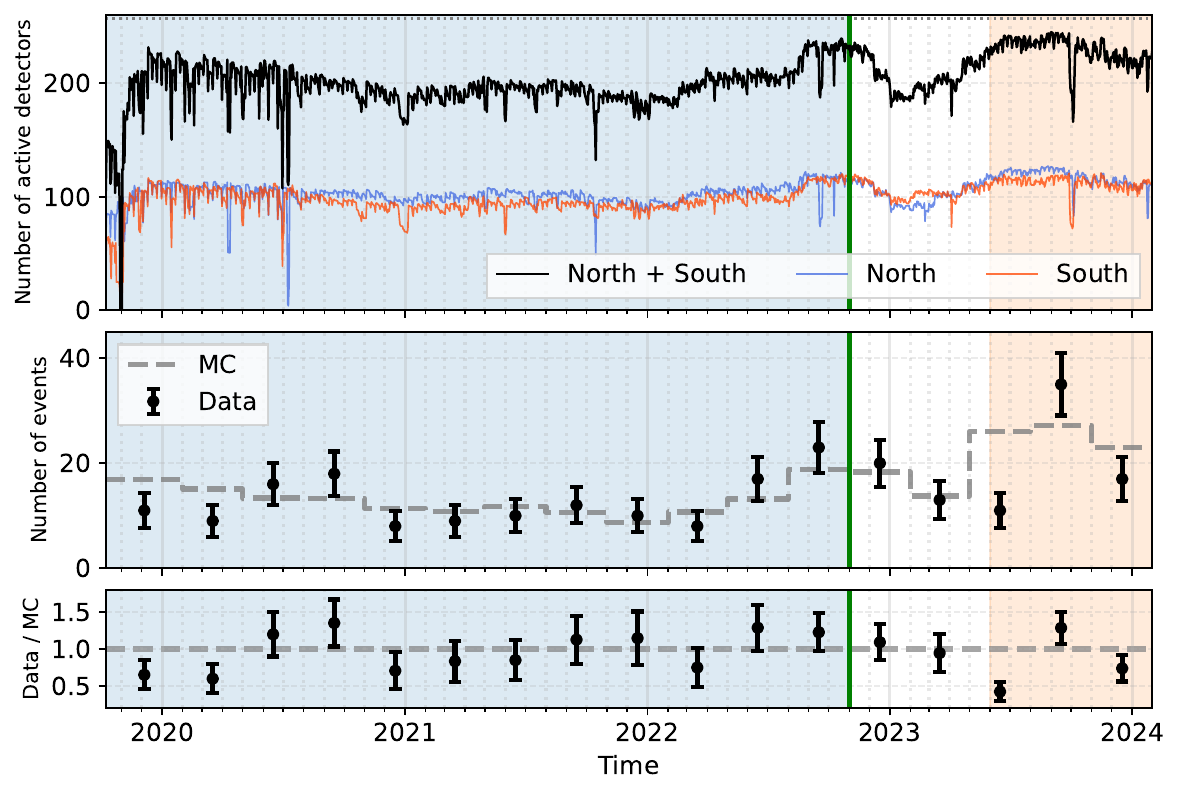}
\caption{(Top) The daily number of active SDs in the TA$\times$4 northern SD array (blue), southern SD array (red), and the sum (black) for the observation period from 8 October 2019 to 31 January 2024. 
(Middle) The number of observed events per time bin after applying the event selection (black points with statistical error bars) and that predicted by the MC simulation (dashed histogram). 
(Bottom) The ratio of the observed number of events to the MC prediction. 
The vertical dark green line represents the time when the inter-tower trigger was implemented. 
The shaded regions in blue and orange indicate the periods used for calculating the effective aperture, shown in Fig.~\ref{fig: eff_aperture_comparisons}.}
\label{fig:occupancy}
\end{center}
\end{figure}

Figure~\ref{fig:occupancy} shows the number of active SDs and the number of reliably reconstructed events observed by the TA$\times$4 SD arrays between October 2019 and the beginning of 2024. 
As seen in the figure, the number of active SDs changed twice: in August 2022 and again in December 2022. 
The first corresponds to the result of the first major on-site maintenance campaign after the relaxation of exceptional international travel restrictions from the COVID-19 pandemic. 
The second, a decrease in the active SDs, resulted from heavy snowfall at the observation site. 
The persistent overcast conditions and low temperatures damaged the batteries and electronics. 
Many batteries were replaced after January 2023, and electronics were replaced between May and June. 
As a consequence, the fraction of active SDs has been greater
than approximately 90\% since June 2023.

In addition to these changes in detector status, the data-taking period is divided into two distinct intervals based on the trigger configuration. 
On 31 October 2022, the trigger system was upgraded by implementing the inter-tower trigger described in Sec.~\ref{sec: Trigger system}. 
This upgrade interconnected the three sub-arrays, which had previously operated independently. 
In Fig.~\ref{fig:occupancy}, this timing is indicated by a vertical green line, marking the transition from the independent sub-array operation to the integrated array operation.

To validate the simulation framework against these dynamic operational conditions, 
the status of each SD was incorporated into the Monte Carlo (MC) simulation (details in Sec.~\ref{sec:MC}). 
To quantitatively evaluate the agreement between the data and the simulation, 
a $\chi^{2}$ test was performed on the event counts per time bin shown in the middle panel of Fig.~\ref{fig:occupancy}. 
The test yields $\chi^{2}/\mathrm{ndof}=24.05/17$ with a p-value of $0.118$. 
Although visible deviations are present in specific intervals, such as early summer 2023, 
the result demonstrates that there is no statistically significant discrepancy over the entire observation period, 
confirming that the MC simulation well reproduces the observed event rate and validating the simulation's reliability.

\section{SD array performance}
\label{sec:performance}
In this section, we report the performance — energy resolution, angular resolution, and aperture — of the TA$\times$4 SD array of the initial 4.3 years, 
after first introducing the MC simulation for the SD array. 
The calibrations described in Sec.~\ref{sec:Lowleveldata} and the status monitoring in Sec.~\ref{sec:workingstatus} are taken into account in the simulation, and thus the derived performance. 
As tested in Sec.~\ref{sec:workingstatus}, the MC simulation reproduces the observed event rate over the observation period, 
validating the simulation's use for the performance evaluation presented below.
\subsection{Monte Carlo simulation for the TA$\times$4 SD}
\label{sec:MC}
We use Monte Carlo (MC) simulation to assess the performance of the TA$\times$4 SD array. 
The details of the simulation used in this study are the same as those described in \citep{Fujisue:2023hkg}. 
The simulation procedure is as follows: 
(i) air showers are generated using CORSIKA \citep{Heck:1998vt} (version 7.3500) with the thinning method described in \citep{Kobal:2001jx} to reduce calculation time. 
QGSJET II-04 \citep{Ostapchenko:2013pia} and FLUKA \citep{Ferrari:2005zk} are used as hadronic interaction models for high- and low-energy regions, respectively. 
EGS4 \citep{Nelson:1985ec} is used for electromagnetic interactions. 
We generated air showers from 10$^{17.5}$ eV to 10$^{20.5}$ eV, with 200 events in each $\Delta$log$_{10}(E/\text{eV})=0.1$ bin. 
The injected zenith angle ($\theta$) follows the $\sin\theta\cos\theta$ distribution, 
which reflects the flat-array detector geometry under the assumption of isotropic cosmic ray arrival, 
and is sampled up to 60$^{\circ}$.
In this study, the injected primary particles are all protons. 
(ii) The secondary particles reaching the ground level are divided into square tiles of 6-m sides, and 20-ns time bins; 
this time binning corresponds to the sampling interval of the surface detector electronics, which is determined by the clock frequency supplied to the FADC.  
A virtual SD is placed at the center of each square tile, and energy deposit in the SD is calculated by a detector response table developed using the GEANT4 \citep{GEANT4:2002zbu} package. 
In this step, the de-thinning procedure described in \citep{STOKES2012759} is applied, which restores statistical information lost during the thinning process. 
(iii) By selecting appropriate tiles, the configuration of an SD array for an air shower with a certain azimuthal angle and a core position is constructed in the simulation. 
In production mode, we reshuffle the generated tiles to simulate different azimuthal angles and core positions.
At this stage, each simulated event is also assigned a specific timestamp, independently of the air shower generation itself. 
The calibration data, described in Sec.~\ref{sec:calibdata}, corresponding to that assigned time are then applied to each SD to reproduce the actual operational status of the SDs.
(iv) The deposited energy is converted to FADC counts by simulating PMT and electronics responses following the calibration data. 
(v) Trigger simulation is applied according to the criteria described in Sec.~\ref{sec: Trigger system} to create a set of events in the same data format as the real data. 
These simulated events are then reconstructed using the exact same pipeline as the data analysis. 
Finally, we weight the simulated events according to the cosmic ray flux measured by the TA experiment \citep{Ivanov:2020rqn} in proportion to the actual exposure of the TA$\times$4 SD array, 
allowing for a direct comparison with the observed data.

It should be noted that this baseline study utilizes pure proton primaries to establish the core performance of the expanded array. 
While the absolute energy scale is ultimately anchored to the FD measurements, a different primary composition, such as iron, can affect shower development and the particle distributions at ground level, 
and consequently modify the MC-based $S_{800}$-to-energy relation, trigger efficiency, and energy resolution. 
A comprehensive composition-dependent evaluation utilizing alternative primary models is deferred to future physics analyses.

\subsection{Event reconstruction and event selection}
\label{sec: Rec & EventSelection}
\begin{figure}[htbp]
    \begin{tabular}{cc}
      \begin{minipage}[t]{0.5\hsize}
        \centering
        \includegraphics[keepaspectratio, scale=0.41]{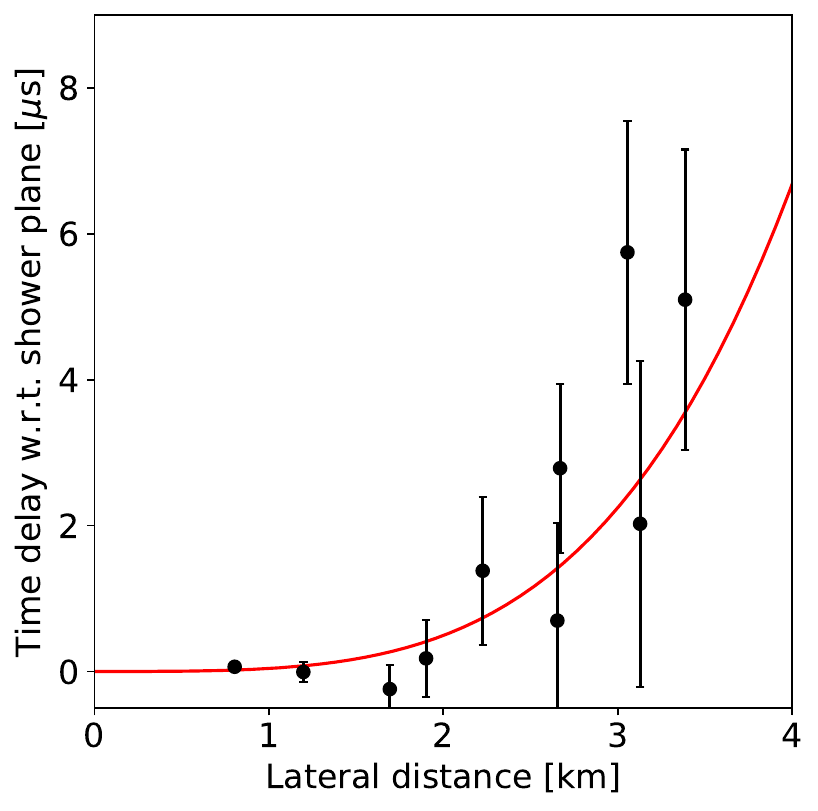}
      \end{minipage} &
      \begin{minipage}[t]{0.5\hsize}
        \centering
        \includegraphics[keepaspectratio, scale=0.41]{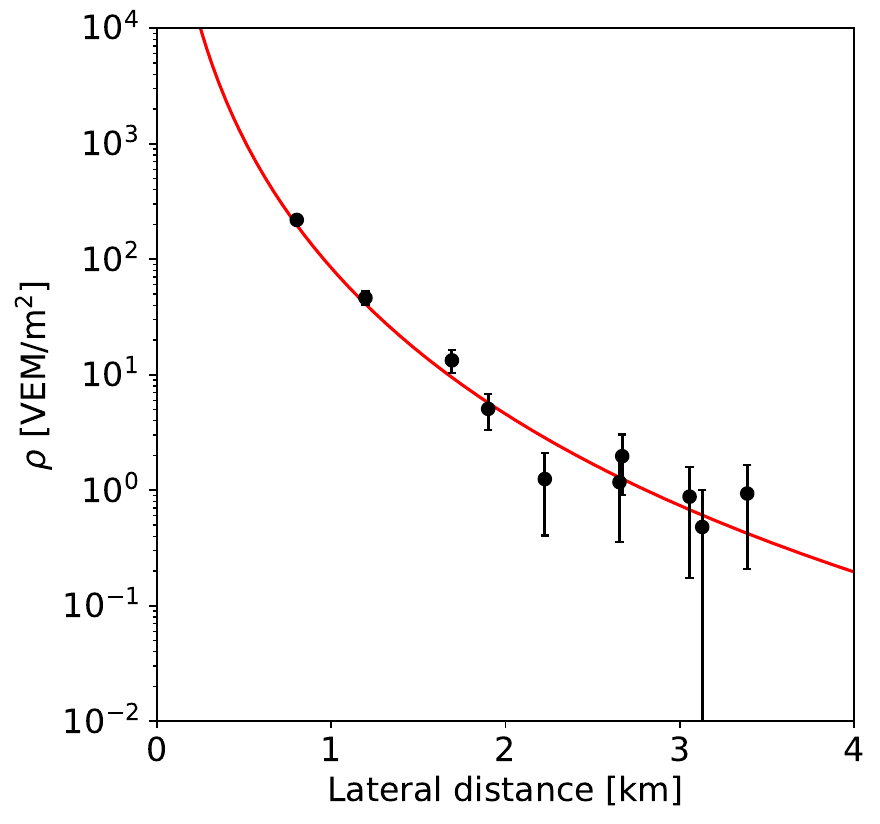}
      \end{minipage} 
    \end{tabular}
    \caption{Examples of the time fit (left) and the lateral distribution fit (right) for a TA$\times$4 SD array event. 
    The event is the same one displayed in Fig.~\ref{fig: event-waveforms}. 
    In both panels, the horizontal axes show the perpendicular distance from each SD to the shower axis. 
    In the left panel, the black points represent the measured time delay from the shower plane, 
    while in the right panel they represent the measured signal density at each SD.
    }
     \label{fig: LDF-ex}
  \end{figure}

To establish the baseline reconstruction framework for the TA$\times$4 SD array, we adopt the same fitting functions used for the original TA SD array \cite{TelescopeArray:2012qqu}.
The arrival direction and the shower core position are determined by fitting the signal timing with the modified Linsley function \cite{Teshima:1986rq}, 
and the profile of the air shower particle density is determined by fitting the signal size to the lateral distribution function as modified by the AGASA experiment \cite{Yoshida:1994jf}. 
The energy of the primary particle is estimated from reconstructed zenith angle and extrapolated $S_{800}$, which is the air shower particle density evaluated at a perpendicular distance of 800 m from the shower axis. 
Using the MC simulations described in Sec.~\ref{sec:MC}, we developed a conversion function that maps the reconstructed zenith angle and $S_{800}$ to the primary particle energy for the TA$\times$4 SD array.
An example of the fit results is shown in Fig.~\ref{fig: LDF-ex}.

The energy estimated by the SD array may have biases stemming from the hadronic interaction model and the detector spacing of the SD array. 
For the original TA SD array, these intrinsic biases were corrected by scaling to match the energy reconstructed from FD signals in hybrid events, since the FD provides nearly a calorimetric measurement of the primary cosmic-ray energy. 
In particular, $E_{\text{TA SD}}/1.27$ is used as a reconstructed energy for the TA SD array, 
where $E_{\text{TA SD}}$ is the energy determined by the conversion function for the TA SD array and the scaling factor of 1/1.27 is determined by direct comparison to the energy from the FD signals \citep{TelescopeArray:2012qqu}.

For the TA$\times$4 SD array, however, we do not yet have enough data to determine the energy scale because the efficiency of the TA$\times$4 SD array decreases quickly below 10$^{19.5}$ eV, and the number of events simultaneously observed by FD and SD is not yet sufficient to render an accurate scaling. 
To overcome this limitation in the current study, we determine a preliminary energy scale factor by normalizing the integral event rate above $10^{19}$ eV to match the cosmic-ray flux measured by the TA SD array \citep{Ivanov:2020rqn}. 
This procedure yields a scaling factor of $1 / (1.36 \pm 0.05 \text{ (stat.)})$.
Note that this factor currently accounts only for statistical uncertainty. 
Systematic uncertainties, such as those arising from the choice of the primary cosmic-ray spectrum model and the energy threshold selection used to derive the factor, have not yet been evaluated.
Given this approach, the reconstructed energy for the TA$\times$4 SD array is provisionally defined as $E_{\text{rec}}=E_{\text{TA}\times\text{4 SD}}/1.36$ in this study, where $E_{\text{TA}\times\text{4\, SD}}$ is the energy determined by the conversion function derived from the TA$\times$4 simulations.
To evaluate the performance on this consistent basis, the true primary energy generated in the MC simulation is also scaled by this 1/1.36 factor, which we hereafter denote as $E_{\text{gen}}$.

This preliminary scaling factor, $1 / (1.36 \pm 0.05 \text{ (stat.)})$, exhibits a slight difference from that of the original TA SD array (1/1.27). 
This discrepancy is qualitatively attributed to the difference in the detector array geometry; 
the TA$\times$4 array has a wider spacing (2.08 km) compared to the original TA array (1.2 km). 
The larger spacing inherently reduces the sampling of the shower core and increases the relative weight of the peripheral detectors in the lateral distribution fitting.
Consequently, this geometric shift in the sampled region could introduce a systematic bias in the energy estimation.

The reconstructed events that pass the following seven conditions are used in the subsequent analyses: 
(1) at least five SDs are used in the reconstruction, 
(2) the reconstructed core position is located at least 400 m inside the array boundary, 
(3) the reconstructed zenith angle is less than 55$^{\circ}$, 
(4) the $\chi^2$/ndof of the reconstruction fit is less than four, 
(5) the uncertainty of the reconstructed direction is less than $6^{\circ}$, 
(6) the fractional uncertainty of the $S_{800}$ is less than 0.5, and 
(7) the reconstructed energy $E_{\text{rec}}$ is greater than 10$^{19}$ eV.

\subsection{Energy and angular resolutions}
\label{sec:Resolution}

\begin{figure}[htbp]
    \begin{tabular}{ccc}
      \begin{minipage}[t]{0.47\hsize}
        \centering
        \includegraphics[keepaspectratio, scale=0.3]{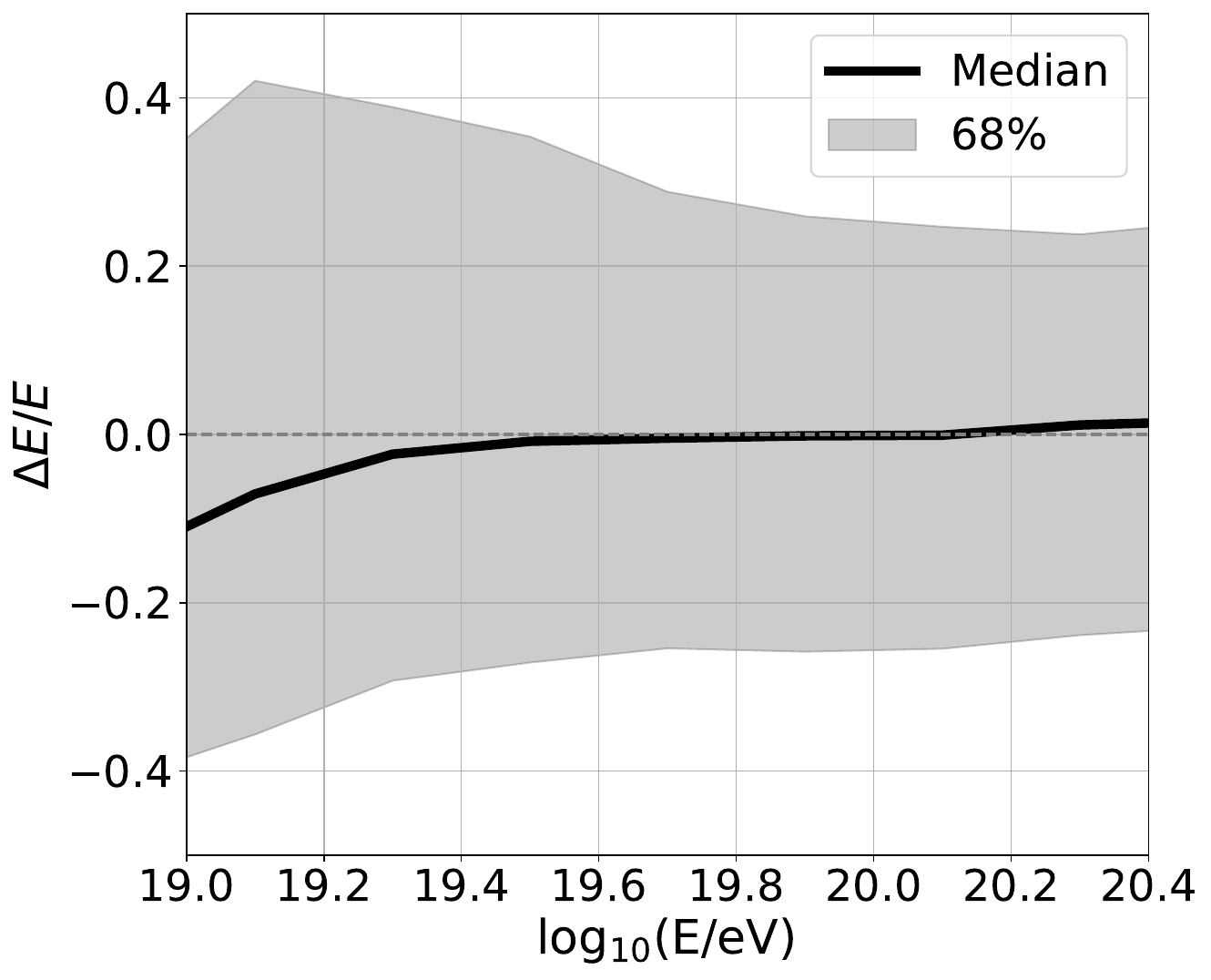}
      \end{minipage} &
      \begin{minipage}[t]{0.47\hsize}
        \centering
        \includegraphics[keepaspectratio, scale=0.3]{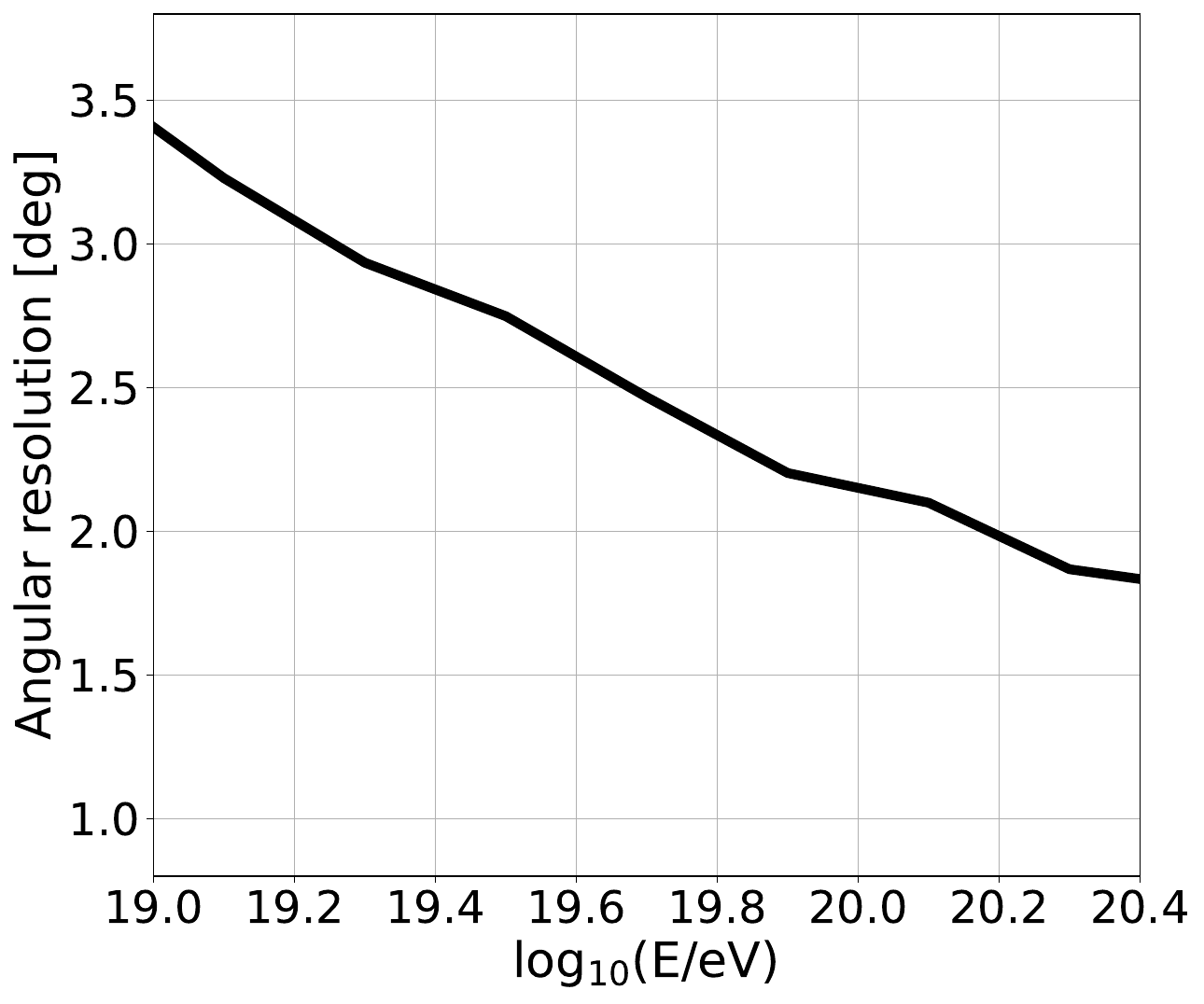}
      \end{minipage} 
    \end{tabular}
    \caption{Reconstruction resolutions for the cosmic-ray energy (left) and arrival direction (right). 
    In the left panel, the median and the 68\% region are indicated by the solid line and the band, respectively. 
    In the right panel, the solid line represents the 68th percentile of the opening-angle between the generated and reconstructed arrival directions.}
     \label{fig: E reso bias}
\end{figure}

We evaluate energy resolution using the distribution of $(E_{\text{rec}} - E_{\text{gen}}) / E_{\text{gen}}$ for the simulated events. 
As defined in Sec.~\ref{sec: Rec & EventSelection}, both  $E_{\text{rec}}$ and $E_{\text{gen}}$ are consistently scaled by the 1.36 factor, ensuring a fair evaluation on the provisional energy scale. 
The angular resolution is evaluated by calculating the 68th percentile of the cumulative distribution of the opening angle $\delta_{\text{ang}} = \arccos (\hat{\textbf{n}}_{\text{rec}}\cdot\hat{\textbf{n}}_{\text{gen}})$, 
where $\hat{\textbf{n}}_{\text{rec}}$ and $\hat{\textbf{n}}_{\text{gen}}$ are the reconstructed and generated arrival direction of an event, respectively. 
As shown in Fig.~\ref{fig: E reso bias}, the overall performance improves with increasing energy. 
Specifically, at $10^{20}$ eV, the energy and angular resolutions are approximately 25\% and 2.2$^{\circ}$, respectively.
The energy and angular resolutions are approximately identical across periods with different trigger systems under the same event selection described in Sec.~\ref{sec: Rec & EventSelection}.

\subsection{Aperture and exposure}
\label{sec:Aperture}
\begin{figure}[!htbp]
\begin{center}
\includegraphics[width=100mm]{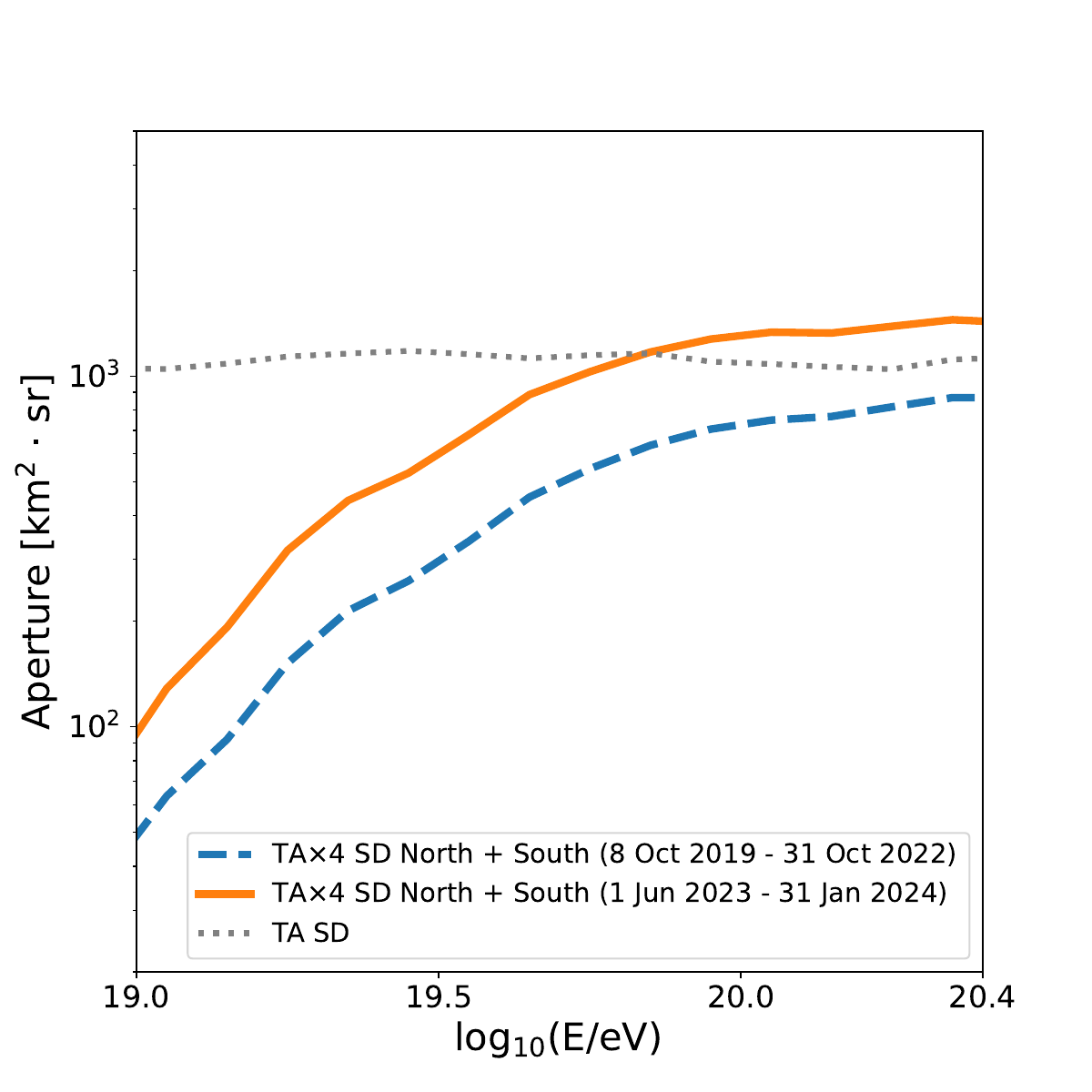}
\caption{Effective apertures of the TA$\times$4 SD array (North + South) before the implementation of the inter-tower trigger, averaging period from 8 October 2019 to 31 October 2022 (blue dashed line), 
and since 1 June 2023 when more than approximately 90\% of the deployed TA$\times$4 SDs have been operating with the inter-tower trigger (orange solid line). 
The aperture of the TA SD array for zenith angles up to 55$^{\circ}$ is also indicated with a gray dotted line.}
\label{fig: eff_aperture_comparisons}
\end{center}
\end{figure}

We evaluate the energy-dependent aperture, $\alpha(E)$, for each energy bin using the MC simulation as follows:
\begin{equation}
\
\alpha(E) = A \cdot \Omega \cdot \frac{N_{\text{sel}}(E)}{N_{\text{gen}}(E)},
\end{equation}
where $A$ and $\Omega$ are the total injection area and the solid angle used in the MC generation, respectively. 
$N_{\text{gen}}(E)$ is the number of generated events in a given energy bin $E$ (which corresponds to the scaled energy $E_{\text{gen}}$), 
and $N_{\text{sel}}(E)$ is the number of events that survive the trigger and the event selection described in Sec.~\ref{sec: Rec & EventSelection}.
Although the aperture near the threshold exhibits a zenith angle dependence, where more inclined showers show a higher trigger efficiency, 
the zenith-integrated aperture is employed here by assuming an isotropic cosmic-ray flux,
following the standard procedure for the energy spectrum determination.
Based on the time-dependent detector status and trigger conditions implemented in the MC simulation, the apertures of the TA$\times$4 SD array, 
which combine all 257 SDs and cover an observational area of 1,000 km$^2$, are evaluated separately for the two observational periods and shown in Fig.~\ref{fig: eff_aperture_comparisons}. 
Since June 2023 (solid orange line in the figure), when over 90\% of the TA$\times$4 SDs have been operating with the inter-tower trigger, the aperture of the TA$\times$4 SD array is 1,300 km$^2$~sr at $10^{20}$ eV, 
approximately 1.2 times that of the TA SD array. 
The introduction of the inter-tower trigger enhances the aperture by approximately 50\%. 
This enhancement of aperture is evaluated by simulating the aperture with and without the inter-tower trigger system for the same observational period. 
The increase of the aperture due to the introduction of the inter-tower trigger is approximately equal to the contribution of the boundary region to the observational area when considering the area excluding 1,000 m (0.48 detector spacing) from the edge of the array.

\begin{figure}[!htbp]
\begin{center}
\includegraphics[width=100mm]{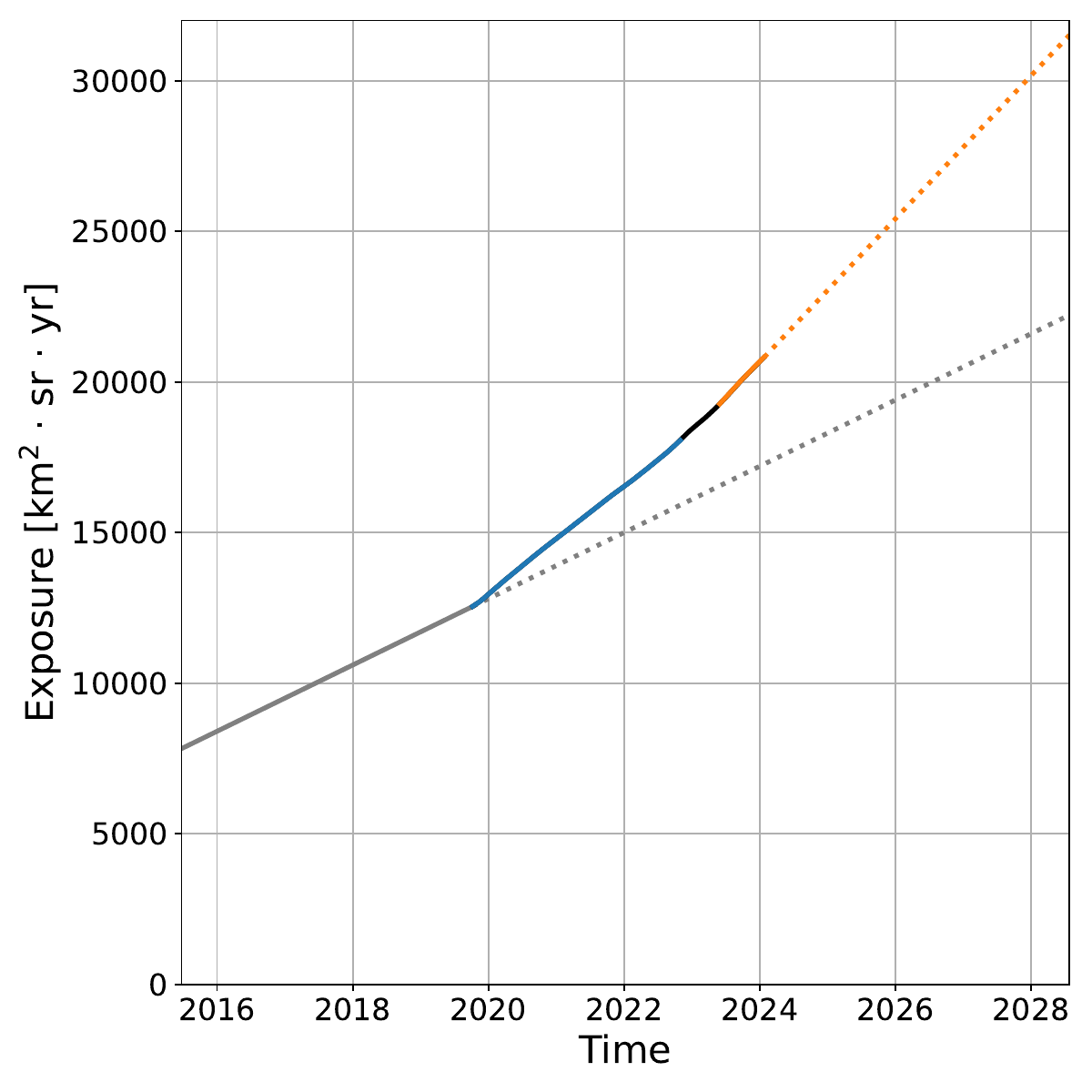}
\caption{Integrated exposure of the sum of the TA$\times$4 northern and southern arrays and the TA SD array (solid line) for zenith angles up to 55$^{\circ}$ at $10^{20}$ eV. 
The blue part indicates the period from 8 October 2019 to 31 October 2022. 
The orange part indicates the period from 1 June 2023 to 31 January 2024.
These periods correspond to the periods shown in Fig.~\ref{fig: eff_aperture_comparisons}. 
The dotted lines represent extrapolation of the exposure: gray for the TA SD array and orange for the total exposure of the half-expanded TA$\times$4 array, assuming a constant average aperture since 1 June 2023, and the TA SD array.}
\label{fig:Exposure}
\end{center}
\end{figure}

We calculate the integrated exposure, $\epsilon(E)$, by integrating period-specific apertures over the total observation time:
\begin{equation}
\epsilon(E) = \int \alpha(E, t) \, dt \approx \sum_{i} \alpha_{i}(E) \cdot \Delta t_{i},
\end{equation}
where $\alpha_{i}(E)$ represents the aperture evaluated under the specific trigger and operational conditions of the $i$-th period with a duration of $\Delta t_{i}$. 
Figure~\ref{fig:Exposure} shows the integrated exposure as a function of time for the combination of the original TA SD array and the extended SD array (TA$\times$4 northern and southern arrays) at $10^{20}$ eV. 
During its first 4.3 years of operation, the extended SD array achieved an exposure of 3,500 km$^2$~sr~yr at $10^{20}$ eV, which corresponds to 20\% of the TA SD array exposure over 15.8 years.

\section{Summary}
\label{sec:summary}
The TA$\times$4 experiment is designed to increase the data acquisition rate of UHECRs by expanding the observation area. 
The upgrade, which started operating in 2019, aims to unveil the origins of UHECRs. 
The total observation area of the existing 257 TA$\times$4 SDs and the original TA SD array is about 2.5 times that of the original TA SD array. 
We analyzed the data acquisition status and the performance of the expanded TA$\times$4 SD array with 257 SDs in the initial period. 
The inter-tower trigger system was implemented in November 2022 to enable triggering in the inter-sub-array region by linking the communication towers, resulting in an approximate 50\% improvement in aperture. 
The number of active SDs also affects the aperture. 
After the relaxation of the travel restrictions due to the COVID-19 pandemic, 
we performed extensive onsite repair and maintenance, and more than 90\% of the deployed SDs have been active since June 2023. 
The energy and angular resolutions at $10^{20}$ eV are approximately 25\% and 2.2$^{\circ}$, respectively. 
The aperture at $10^{20}$ eV after the implementation of the inter-tower trigger is 1,300 km$^2$~sr, approximately 1.2 times that of the TA SD array. 
The integrated exposure of the TA$\times$4 SD array for the first 4.3 years is 3,500 km$^2$~sr~yr at $10^{20}$ eV, representing 20\% of the integrated exposure of the TA SD for the 15.8 years of observation. 
The TA$\times$4 SD array achieves the performance expected with the MC simulation. 
Keeping the stable operation of the TA$\times$4 SD array will enhance the number of UHECR events in the northern hemisphere, which is crucial for elucidating the origins of UHECRs. 
By completing the full TA$\times$4 SD array, we will increase the observational area by another 1,000 km$^2$, and further accelerate the overall rate of data collection.

\section*{Acknowledgments}
The Telescope Array experiment is supported by the Japan Society for
the Promotion of Science (JSPS) through
Grants-in-Aid
for Priority Area
431,
for Specially Promoted Research
JP21000002,
for Scientific  Research (S)
JP19104006,
for Specially Promoted Research
JP15H05693,
for Scientific  Research (S)
JP19H05607,
for Scientific  Research (S)
JP15H05741,
for Science Research (A)
JP18H03705,
for Young Scientists (A)
JPH26707011,
for Transformative Research Areas (A)
JP25H01294,
for International Collaborative Research
24KK0064,
and for Fostering Joint International Research (B)
JP19KK0074,
by the joint research program of the Institute for Cosmic Ray Research (ICRR), The University of Tokyo;
by the Pioneering Program of RIKEN for the Evolution of Matter in the Universe (r-EMU);
by the U.S. National Science Foundation awards
PHY-1806797, PHY-2012934, PHY-2112904, PHY-2209583, PHY-2209584, and PHY-2310163, as well as AGS-1613260, AGS-1844306, and AGS-2112709;
by the National Research Foundation of Korea
(2017K1A4A3015188, 2020R1A2C1008230, and RS-2025-00556637);
by the Ministry of Science and Higher Education of the Russian Federation under the contract 075-15-2024-541, 
IISN project No. 4.4501.18, by the Belgian Science Policy under IUAP VII/37 (ULB), 
by National Science Centre in Poland grant 2020/37/B/ST9/01821, 
by the European Union and Czech Ministry of Education, 
Youth and Sports through the FORTE project No. CZ.02.01.01/00/22\_008/0004632, and by the Simons Foundation (MP-SCMPS-00001470, NG). 
This work was partially supported by the grants of the joint research program of the Institute for Space-Earth Environmental Research, 
Nagoya University and Inter-University Research Program of the Institute for Cosmic Ray Research of University of Tokyo. 
The foundations of Dr. Ezekiel R. and Edna Wattis Dumke, Willard L. Eccles, and George S. and Dolores Dor\'e Eccles all helped with generous donations. 
The State of Utah supported the project through its Economic Development Board, and the University of Utah through the Office of the Vice President for Research. 
The experimental site became available through the cooperation of the Utah School and Institutional Trust Lands Administration (SITLA), U.S. Bureau of Land Management (BLM), and the U.S. Air Force. 
We appreciate the assistance of the State of Utah and Fillmore offices of the BLM in crafting the Plan of Development for the site.  We thank Patrick A.~Shea who assisted the collaboration with much valuable advice and provided support for the collaboration’s efforts. 
The people and the officials of Millard County, Utah have been a source of steadfast and warm support for our work which we greatly appreciate. 
We are indebted to the Millard County Road Department for their efforts to maintain and clear the roads which get us to our sites. 
We gratefully acknowledge the contribution from the technical staffs of our home institutions. 
An allocation of computing resources from the Center for High Performance Computing at the University of Utah as well as the Academia Sinica Grid Computing Center (ASGC) is gratefully acknowledged.
E.K. acknowledges support from the Chinese Academy of Sciences President’s International Fellowship Initiative (Grant No. 2025PVC0021) and from JSPS KAKENHI Grant No. JP25K07333.
K.F. acknowledges support from Academia Sinica (Grant No. AS-GCS-113-M04) and the National Science and Technology Council (Grant No. 113-2112-M-001-060-MY3).







\end{document}